%% file: apc-proceedings.tex
\documentclass[a4paper]{jpconf}
\usepackage{amsmath}
\usepackage{url}
\usepackage{graphicx}
\usepackage{caption}
\usepackage{subcaption}
\usepackage{enumitem} 

\setlist[itemize]{noitemsep}

\begin{document}
\title{How do particle physicists learn the programming concepts they need?}

\author{ S Kluth$^1$, M G Pia$^2$, T Schoerner-Sadenius$^3$, P Steinbach$^4$}

\address{$^1$ Max Planck Institute for Physics, F\"ohringer Ring 6, 80805 Munich, Germany}

\address{$^2$INFN Sezione di Genova, Via Dodecaneso 33, 16146 Genova, Italy }

\address{$^3$ DESY-FH/CMS, Notkestr. 85, D-22607 Hamburg, Germany}

\address{$^4$ Scientific Computing Facility, Max Planck Institute of Molecular Cell Biology and Genetics, Pfotenhauerstr. 108, 01307 Dresden, Germany }

\ead{Maria.Grazia.Pia@cern.ch, skluth@mpp.mpg.de, thomas.schoerner@desy.de, steinbac@mpi-cbg.de}

\begin{abstract}
The ability to read, use and develop code efficiently and successfully is a key
ingredient in modern particle physics.
%Software design plays a fundamental role in the software development process and is instrumental to many critical aspects in the life-cycle of an experiment: the transparency of software design enables the validation of physics results, contributes to the effective use of human and computational resources, and facilitates the evolution and the maintainability of the software over the lifetime of an experiment. 
We report the experience of a training program, identified as ``Advanced
Programming Concepts'', that introduces software concepts, methods and
techniques to work effectively on a daily basis in a HEP experiment or other
programming intensive fields.
This paper illustrates the principles, motivations and methods that shape the
``Advanced Computing Concepts'' training program, the knowledge base that it
conveys, an analysis of the feedback received so far, and the integration of
these concepts in the software development process of the experiments as well as
its applicability to a wider audience.
\end{abstract}

\section{Introduction}
\label{sec:intro}

Large software systems play a fundamental role in detector simulation, detector
operation, data read-out and analysis \cite{Antcheva20092499, Bockelman:2014bea,
%Clemencic:2010zz, Chauhan:2014oga, Allison:2006ve} in modern particle and
Clemencic:2010zz, Chauhan:2014oga, g4nim} in modern particle and
nuclear physics.
The software systems used hereby involve object-oriented (OO) frameworks of
typically more than $10^{5}$ lines of source code.
As the task they try to tackle exposes a high complexity, the software that maps
a solution of the task is complex as well.

Scientists that would like to record data, build new detectors or analyse data
are required to use these systems to extract knowledge, improve detectors
designs and test scientific hypotheses in order to eventually answer scientific
questions.
This applies to all stages of the academic career: it concerns students,
post-docs and senior physicists.
However, often physicists are not adequately trained by means of the standard
university curriculum to object-oriented programming (OOP).

This has lead to the situation, that a high proportion of the daily academic
work is dedicated to learn how to program, to understand and potentially fix
source code and to layout and implement new functionality that is either missing
or present in an unusable fashion. 
As this fact has to be acknowledged as a
reality, there was a need for suitable training on the topic for physicists.

We report the experience of a training program, identified as ``Advanced
Programming Concepts'' (APC), that introduces software concepts, methods and
techniques to work effectively on a daily basis in a HEP experiment or other
programming intensive fields.
The program is targeted at students and young researchers involved in physics
analysis and detector development or related software heavy activities, not only
at core software developers of relevant scientific code bases.
The APC workshop introduces basic and advanced programming techniques as well as
elements of the software development process and project management skills.
Emphasis is given to methods on how to work effectively with existing code, to
improve code and to build a basis for further self-improvement in the field.

The paper is laid out as follows: section \ref{sec:ideas-behind} identifies the
key motivations behind the APC curriculum, section \ref{sec:contents}
illustrates the contents of the school and provides insight in the pedagogical
ideas applied, section \ref{sec:eval} highlights key findings of the participant
survey that was conducted and finally section \ref{sec:summ} concludes the
discussion.

\section{Ideas behind the School}
\label{sec:ideas-behind}

\input{apc-ideas}

\section{School Layout and Content}
\label{sec:contents}

\subsection{Test-Driven Development and class design principles}
\label{subsec:TDD}

\input{apc-content-steinbac}

\subsection{Object Oriented methods, Unified Modeling Language and Design Patterns}
\label{subsec:DP}

\input{apc-content-skluth}

\subsection{Software Engineering and Refactoring}
\label{subsec:Ref}

\input{apc-content-maria}

\subsection{Performance and modern programming techniques}
\label{subsec:CM}

\input{apc-content-steinbac2}

\section{Evaluation}
\label{sec:eval}

\input{apc-eval}

\section{Summary}
\label{sec:summ}

This report illustrates the principles and methods that shape the "Advanced
Computing Concepts" training program, the knowledge base that it conveys, an
analysis of the feedback received so far, and the integration of these concepts
in the software development process of the experiments as well as its
applicability to a wider audience.

It intends to promote a discussion in the software-oriented particle physics
community on the responsibility of better preparing our young people for their
work in the experiments, and on how the experiments could profit from a wider
knowledge of advanced software methods and techniques.

\ack
%\section*{Acknowledgements}
We would like to thank the Helmholtz Alliance ``Physics at the Terascale'' for
financially supporting the workshop over four years. We also would like to thank
past contributors: Thomas Velz (University of Bonn, now industry) was a
participant once in the workshop and contributed as a teacher in 2014
(\cite{APC-2014}), Benedikt Hegner (CERN, \cite{APC-2011}) and Eckhardt von
Toerne (University of Bonn, \cite{APC-2010}) both made substantial contributions
to past workshops.

\section*{References}

\bibliographystyle{iopart-num}
\bibliography{apc-proceedings}

\appendix
\section*{Appendix}
\subsection{Evaluation Catalog}
\label{app:evaluation}

\begin{enumerate}
\item The year you were born (open text field)
\item Your gender (choice: male or female)
\item Academic position at the time of the workshop (choice: Student, PhD Student, Post-Doc, PI/Group Leader, Technician, Other)
\item My theoretical knowledge on the subject prior to the course (choice: ``1 = Poor'' to ``5 = Excellent'')
\item My practical experience on the subject prior to the course (choice: ``1 = Poor'' to ``5 = Excellent'')
\item Course content (choice: ``1 = Poor'' to ``5 = Excellent'')
\item Course structure (choice: ``1 = Poor'' to ``5 = Excellent'')
\item Preparation of the course (choice: ``1 = Poor'' to ``5 = Excellent'')
\item Focus of the course on relevant points (choice: ``1 = Poor'' to ``5 = Excellent'')
\item Illustration of possible applications (choice: ``1 = Poor'' to ``5 = Excellent'')
\item Course duration appropriate to the content (choice: ``1 = Poor'' to ``5 = Excellent'')
\item Course materials (choice: ``1 = Poor'' to ``5 = Excellent'')
\item Encouragement of the learning process (choice: ``1 = Poor'' to ``5 = Excellent'') 
\item Enthusiasm of lecturer with the subject matter (choice: ``1 = Poor'' to ``5 = Excellent'') 
\item Availability of lecturer for questions during / after course (choice: ``1 = Strongly disagree'' to ``5 = Strongly Agree'')
\item The course is relevant for my current work (choice: ``1 = Strongly disagree'' to ``5 = Strongly Agree'')
\item The course broadened my general comprehension (choice: ``1 = Strongly disagree'' to ``5 = Strongly Agree'')
\item I benefited from the course (choice: ``1 = Strongly disagree'' to ``5 = Strongly Agree'')
\item I enjoyed attending the course (choice: ``1 = Strongly disagree'' to ``5 = Strongly Agree'')
\item I would recommend this course to others (choice: ``1 = Strongly disagree'' to ``5 = Strongly Agree'')
\item General Comments (positive) (open multi-line text field)
\item General Comments (negative) (open multi-line text field)
\end{enumerate}

\end{document}

%% file: apc-ideas.tex
Object-orientation is currently the prevalent programming paradigm adopted in
scientific codes in High Energy Physics (HEP).
However, OOP is hard to learn and even harder to put to efficient and effective
use.

As the generation of scientific results is of utmost priority to most scientific
stakeholders and group leaders, the common approach of handling the complexity
of software for newcomers in research groups is to read-up on internet based
tutorials or arbitrary monographs in order to learn a minimal set of programming
language features.
Then scientists turn to examples or peer produced source code and alter
parameters and individual lines of code to produce the application behavior
aspired.
At the cost of considerable inefficiency, the described mindset is retained and
carried out throughout entire PhD and post-doc careers.
Not only does the task complexity in scientific careers as well as deadline
pressure increase, but also many scientists are expected to contribute back to
experimental code bases which makes these codes subject to software quality and
performance regressions if not approached with care.

Having the above in mind, the APC set out in 2010 (\cite{APC-2010}) to convey
central aspects of OO software design to scientists whose programming skills are
moderate on average.
The core motivation was not to teach the academic depths of multi-facetted
requirement analysis, user story identification and software modeling as
commonly taught in computer science based courses of the like.
We aspired more to extract carpentry-style aspects of software design and map
them to exercises and examples that the participants can relate to.

The idea condensed in teaching recipes to solve re-occurring problems in OOP
which go by the names of ``class design principles'' (\cite{martin2003agile})
and ``design patterns'' (\cite{Gamma:1995:DPE:186897}) and practice these within
well defined exercises together with peers.
We intended to help beginners grasp the OO methods by means of practise, apply and
train proven solutions in a collaborative fashion and most of all take concrete
code snippets back to their home institute.

The APC workshop program is largely different from that of most computing schools 
addressed to young high energy physicists, which focus on teaching elementary C++
programming or are addressed to core developers of software frameworks.
The APC program focuses on the needs of ``normal'' young physicists in their 
everyday work in high energy physics experiments: by far, this category represents 
the vast majority of young physicists. 
It is worthwhile to note that not only individuals, but also the experiments as a whole, 
would benefit from improving the knowledge base, software development skills 
and methodological awareness of this large category of the experiments' manpower.

Four APC workshops were held between 2010 and 2014, which were attended by 138 participants in total.
As the curriculum has evolved over the years, the core ideas of communicating
essential concepts of OOP, training on software design and maintenance remained at the
heart of the school.
Thus, the next section will discuss these in detail with a focus on the workshop
as given in 2014 (\cite{APC-2014}).

%% file: apc-content-steinbac.tex
Many of our participants work in a code-centric environment. 
This means, their programming skills have not yet reached a level of
maturity where they could easily concentrate on the design of the software rather than the
implementation.

Therefore, we started in 2011 to introduce them to unit tests and test-driven
development (TDD, \cite{beck:tdd}).
The idea behind this was to provide a method to assure the programmer that
behavior already implemented does not change due to redesigns or extensions of
the code base.
At the heart of test-driven development, this idea is taken even further whereas
the canonical test-aided programming work-flow (write tests on implemented
functionality) is inverted: the tests of functionality are provided first, then
the implementation satisfying these tests are put in place and eventually the
software design is re-evaluated and/or updated before new tests are provided,
i.e. the TDD cycle starts over.

These concepts are first demonstrated live while implementing a simple vector
class in C++ or python - the choice of language depends on where the majority of
the audience feels most at home in.
Simple member functions such as \texttt{get}, \texttt{set} methods or
\texttt{add} functionality are used as an example to demonstrate the TDD
work-flow based on a xUnit style unit test framework (see SUnit as the first of
this kind, \cite{beck1999kent}) - we used {\tt Boost.Test} (\cite{boosttest})
here.

The students are then asked to practice TDD on their own by extending the vector
to offer a {\tt magnitude} member function which ultimately should even take
different policies on how to calculate the magnitude (Minkowski metric,
Euclidean metric, etc.).
Already during the live coding, the students are encouraged to code along in
order to have everyone see the lecture contents first hand running on his/her
laptop.

Even though, the students are not expected to pick up TDD as their style of
programming, this lecture provides an entrance to contemplating programming from
a higher level. 
TDD not only touches on software quality and reproducibility, it
also makes participants think about independent feature sets of their classes
and a defensive mindset when implementing new features.

The latter discussion of independent feature sets of an object-oriented class,
imminently leads to the question, if there are best practices on how to design a
class.
The class design lecture in APC that tries to answer this follows immediately
and is loosely based on \cite{martin2003agile}.
The Single-Responsibility principle, Open/Closed principle, the Liskov
Substitution principle, Interface Segregation and the Dependency Inversion
principle are discussed in depth.
Also, package design principles are covered if time allows it.

In this part of the workshop, a more abstract level of thinking about OOP is
conveyed and a solid understanding of inheritance and its relation to a
hierarchy of feature sets is achieved.
The Unified Modeling Language (UML) is put to use in order to visualize source code
that complies or refuse the principles states above.
Even though small pen-and-paper exercises to fortify the contents are provided,
the lecture is laid out in an open fashion so as to adapt to the speed of the
participants to a high degree (\cite{ambrose2010learning}).

%% file: apc-content-skluth.tex
Essentially all of our students, and in fact most of the lecturers, did
not have significant formal education during their university studies
on computer science in general or OOP in
particular. 
Therefore students often have difficulties to
appreciate the benefits of object oriented programming for
creating large software systems, since they never had the opportunity to
understand the scientific arguments leading to OOP.
In a lecture based
on classic books on OOP, the fundamental principles are presented and
contrasted with procedural programming in order to highlight the
differences.

The Unified Modelling Language UML is used to
present class relations, e.g.\ to discuss the class design principles.
The APC has a lecture combined with paper-and-pencil exercises to explain the
UML for classes, relations between classes, for objects, and for sequences of
events between objects.
The UML has a detailed formal definition and in the lecture
we only present what is needed to discuss design issues with pencil
and paper or on a blackboard. 
The main idea is to establish a common
language for discussing software design issues which goes beyond
writing down code or pseudo-code. 
The UML concepts are discussed as analogue to developing formulae or Feynman diagrams to
simplify understanding a physics problem and writing down the correct
solution.
UML diagrams are then consistently used to present more advanced
topics in the course of the school. 
With UML a more structured discussion about architecture and design of software
systems becomes possible.

With OOP, class design principles, the UML language and the structured
code development practice of TDD discussed, many students ask with
justification how for a real project they should start.  
At the heart
of this question is that, of course, a single class cannot solve even a
small software project.  
At this point the ``Design patterns'' come into
play~\cite{Gamma:1995:DPE:186897}.  
The design patterns collected in the well-known book were inspired by the idea
of collecting common solutions for common architecure design problems as
initiated by C. Alexander~\cite{calexander}.
In the lecture and exercises the most
important and common object oriented programming class design patterns
are presented and discussed with examples drawn from HEP software whenever
possible. 
Going beyond the GoF (Gang of Four) design patterns some more HEP software
specific design patterns are presented and discussed as well. 
This helps students to grasp the concepts behind many of the large software
systems in HEP.

%% file: apc-content-maria.tex
In a lecture combined with a hands-on session all aspects presented so
far are brought together and combined with the topic of
refactoring~\cite{refac}.  

Refactoring refers to the process of
improving the design of existing code without changing its behaviour and
functionality.  

Dealing with existing code is the situation students and post-doctoral associates 
commonly face in their project assignments in high energy physics experiments,
which are characterized by a long life-cycle.
Since in a young physicist's project either already existing code has to be modified or new code
has to be added and made functional, it is clear that almost always the work
is done inside a body of code which already has some function or behaviour.
The room assigned to refactoring in the APC workshop program recognises this
common situation.
Refactoring is presented in the APC lectures not only as a technique to improve
the design of existing code - often as the a necessary step to modify or add
functionality, but also as a set of practical software design guidelines to
develop the students' new code according to good quality standards.

Emphasis is placed on formalising the development process into a sequence of
small steps, supported by a solid set of tests, where the changes between steps
are small and the tests verify that no errors where introduced and 
the behaviour of the original code was unaltered.

The hands-on session is based on a
prepared small C++ project, which is modified by the students in
several small steps to either change the internal structure to prepare
for adding a new feature, or adding the new feature together with its
unit test.

Only with all three ingredients, namely unit testing, OOP (including for our
purposes class design principles and design patterns) and
refactoring, a complete practice of software engineering applicable to
the daily basic needs of programming in HEP emerges.  
The test discipline guides in building up a
working body of code with built-in verification, OOP (as defined
above) helps in finding an appropriate structure for the code, and
refactoring is the method to systematically make changes to existing
code.  

The conceptual framework of the school drives the students to understand these
relationships and to appreciate how following this systematic approach can make
their activity as programmers in HEP much more productive.

%% file: apc-content-steinbac2.tex
As section \ref{subsec:Ref} covered essential methods and tools to re-engineer
and modernize code, techniques of state-of-the-art C++ and performance
improvements to exploit modern computer architectures have been added to the
curriculum.

HEP based codes yield a characteristic performance footprint: embarassingly parallel 
single node applications heavy on I/O, which exploit data-parallelism on multiple levels. 
For this reason, we introduced one entire half-day lesson that focusses on performance 
measurements and programming techniques to use multi-core systems as well as SIMD 
instructions in a approachable way. 

We put the focus on performance evaluation first by discussing modern 
CPU architectures from a very high level and then turning to open-source 
and free tools to evaluate application performance ({\tt iperf}, {\tt valgrind}, etc.). 
We emphasize the importance of measurements first over the common trend 
to exploit parallelism and related low-level CPU features at will where it might not be needed. 
The latter is a common culprit that many beginners invest too much time in 
without achieving significant gains, and at the same time producing overly complex code. 
For introducing multi-threading into applications, we chose to teach OpenMP (\cite{openmp}). 
For SIMD instructions, we teach compiler based auto-vectorisation (\cite{gcc-sse-av}) and 
SSE intrinsics (\cite{sse-intrinsics}), if time permits.

In recent editions of the APC workshops, we concluded the programme
by returning to a more source code based discussion. 
As the majority of HEP codes apply runtime feature selection by virtual inheritance, 
we introduced a discussion on an alternative way in C++ to structure code. 
For this, we perform a live-coding session without any slides entirely 
(as opposed the lessons discussed in section \ref{subsec:TDD}, 
where live-coding and slide presentation is mixed).
 Here, we start by introducing the \texttt{template} keyword to functions 
in simple C++ applications, going further to templated classes, compile-time interfaces 
as realized by the curiously-recurring-template pattern (\cite{crtp}) 
and finalizing this session by unrolling a loop at compile time.

%% file: apc-eval.tex
Various methods of evaluation are applied to collect feedback from the participants
of the APC workshops. 
Questionnaires circulated at the end of the workshop provide an immediate 
appraisal of the lectures.
For purposes of this paper, we have conducted an evaluation to collect information on long-term effects, 
based on an online survey embedded in each workshops indico web page.
Given the relatively small number of students involved, a significant role is also played by
informal, direct communication between students and lecturers, who remain reachable 
after the formal completion of the school programme.

The evaluation of long-term effects comprised a question catalog of 22 items (see \ref{app:evaluation}). 
Participants were asked to submit their answers anonymously. 
%The repository of this publication at \cite{apc-proceedings-repo} contains the indico configuration file to set up the evaluation (\texttt{data/course-evaluation.xml}) as well as the results from each individual workshop (see \texttt{data/evaluation-results-*csv}).
The evaluation was sent to participants of all workshops. 
In total 23 individuals have replied within the deadline for the submission of this paper; 
thereof 4 participants from the 2010, 4 from 2011, 
5 individuals from 2012 and 11 entries from 2014 were recorded. 
From this sample, a bias towards the workshop program of 2014 is to be expected. 
19 out of 23 participant were PhD students at the time of their course; 4 were Post-Docs.

As the statistics is very limited at the time of writing this paper, 
%the fixed scope of this publication and the comprehensiveness of the question catalog, 
we will only highlight certain aspects of the evaluation. 
Further, all figures listed below were obtained from the full data sets irrespective of the participants' workshop year.

\begin{figure}[hbt]
\centering
  \begin{subfigure}[b]{0.45\textwidth}
    \includegraphics[width=\textwidth]{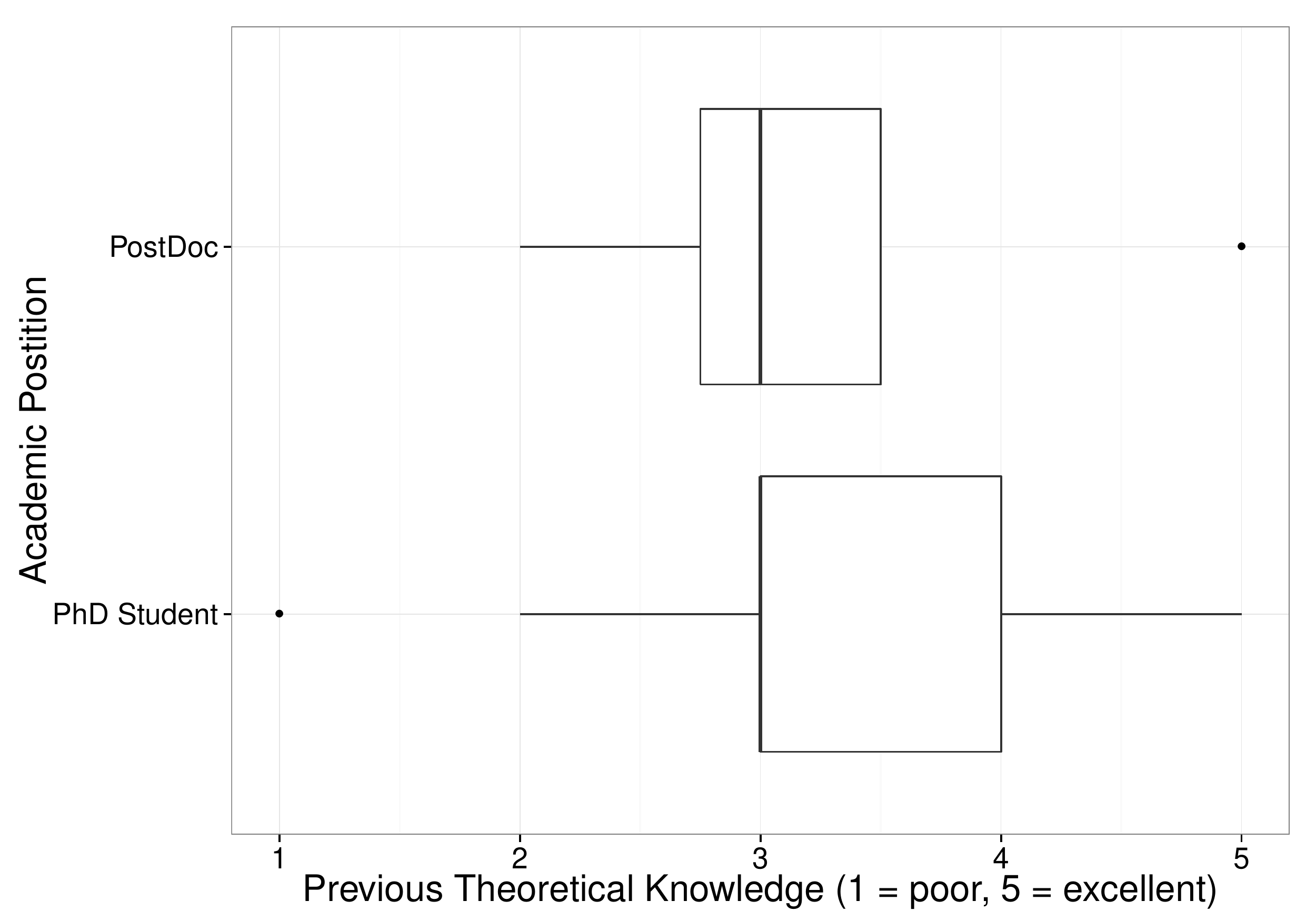}
    \caption{Candidates estimate of their prior theoretical knowledge before the course.}
    \label{sfig:prev_know_theo}
  \end{subfigure}%
  \hfill
  \begin{subfigure}[b]{0.45\textwidth}
    \includegraphics[width=\textwidth]{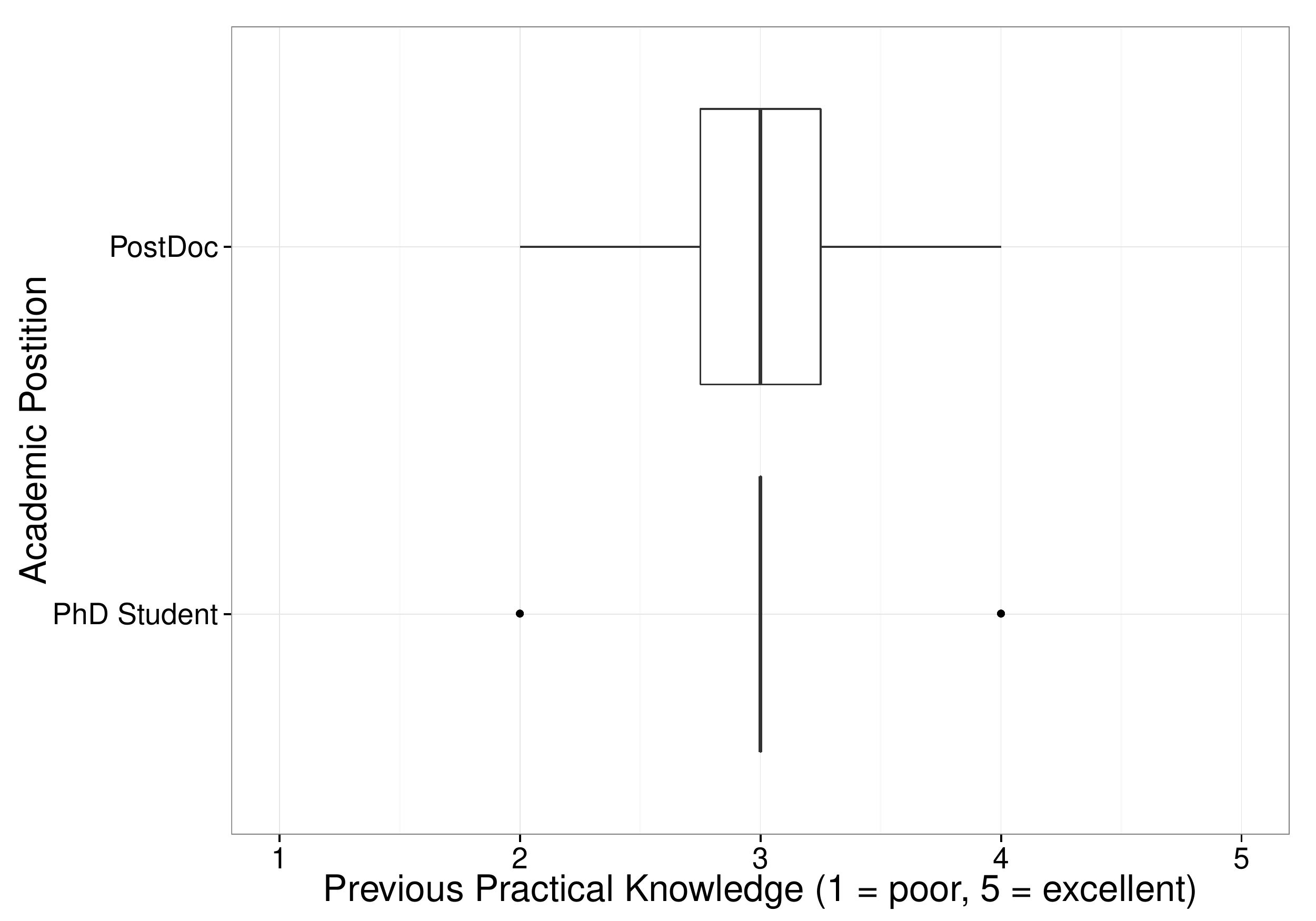}
    \caption{Candidates estimate of their prior practical knowledge before the course.}
    \label{sfig:prev_know_prac}
  \end{subfigure}%

\caption{Box plots of the candidates estimates on their prior knowledge (for sub-figure \ref{sfig:prev_know_theo} referring to theory and for sub-figure \ref{sfig:prev_know_prac} to the practical aspects)  on advanced programming concepts prior to the school. The vertical box outer limits denoted the $25\%$ and $75\%$ quantile limits of the answer. The bold vertical line inside the box marks the arithmetic mean of the sample.}
\label{fig:prev_knowledge}
\end{figure}

Figure \ref{fig:prev_knowledge} summarizes the answers of participants of APC on how they estimated their programming and software design proficiency before the course. 
The evaluation inquired on the theoretical knowledge they had (design patterns, programming methodology, etc.) and on the practical aspects of it (implementation details, language specifics, etc.). 
In both fields, candidates consider themselves average (figure \ref{sfig:prev_know_prac}) with a slight tendency towards good knowledge (figure \ref{sfig:prev_know_theo}). 
%Post-Docs and PhD students share this view on average, although 
%PhD students seem to deem themselves of having good knowledge. 
%The results of figure \ref{fig:prev_knowledge} might be biased, as we asked participants \textit{after} 
%the course about their knowledge \textit{before} course. 
%Under the assumption that participants learned something, 
%this might always tend to too high values. 
%However, the results make clear that the majority do not consider themselves 
%excellent programmers. 
%This observation is potentially due to the fact, that most, if not all, participants 
%are self-taught programmers. 
%With this, they lack the methods and experience
% to judge if their programs or source code they contribute are of high quality or not.

\begin{figure}[htb]
\centering
  \begin{subfigure}[b]{0.45\textwidth}
    \includegraphics[width=\textwidth]{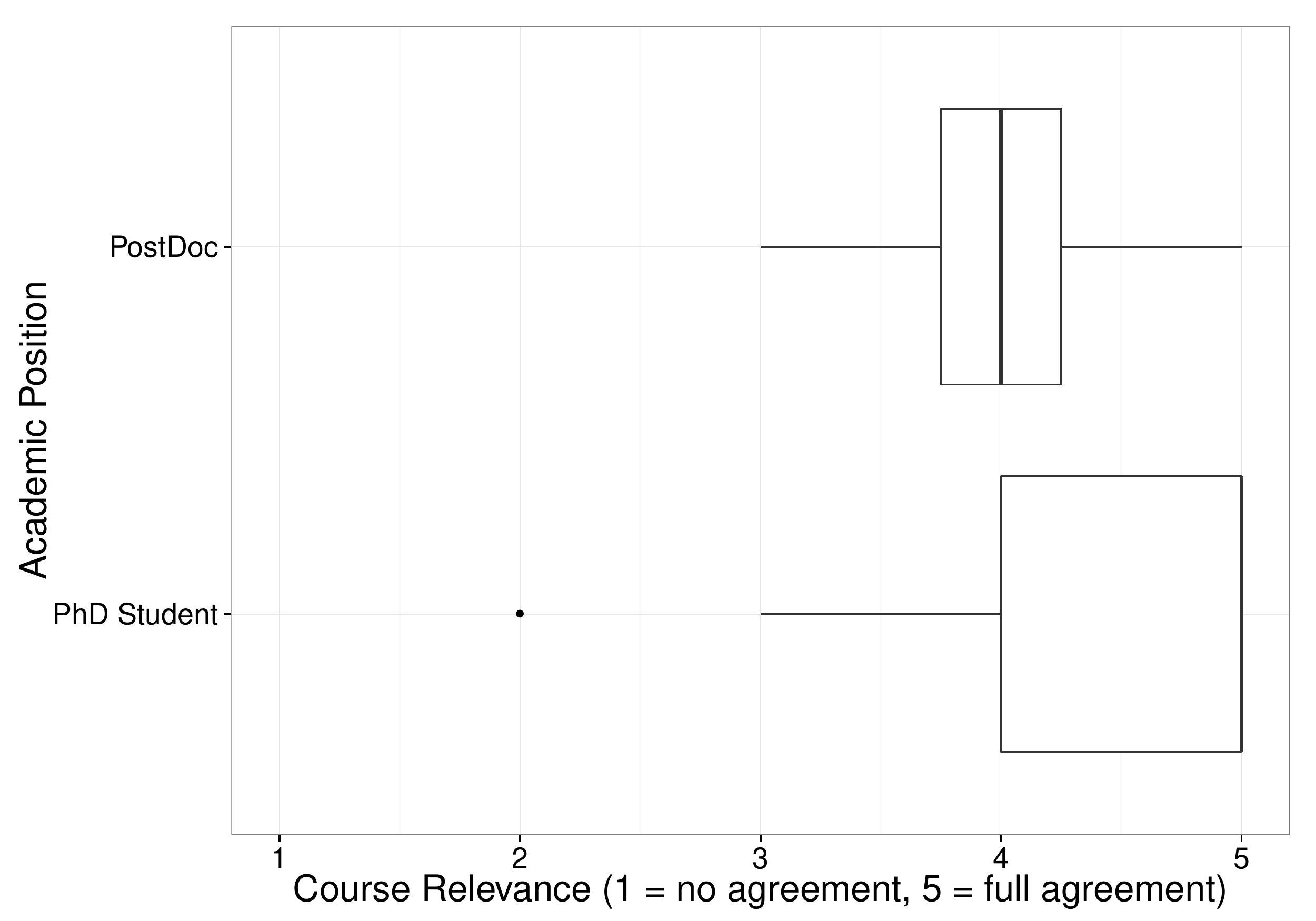}
    \caption{Candidates estimate of the workshops relevance for their work.}
    \label{sfig:ws_relevance}
  \end{subfigure}%
  \hfill
  \begin{subfigure}[b]{0.45\textwidth}
    \includegraphics[width=\textwidth]{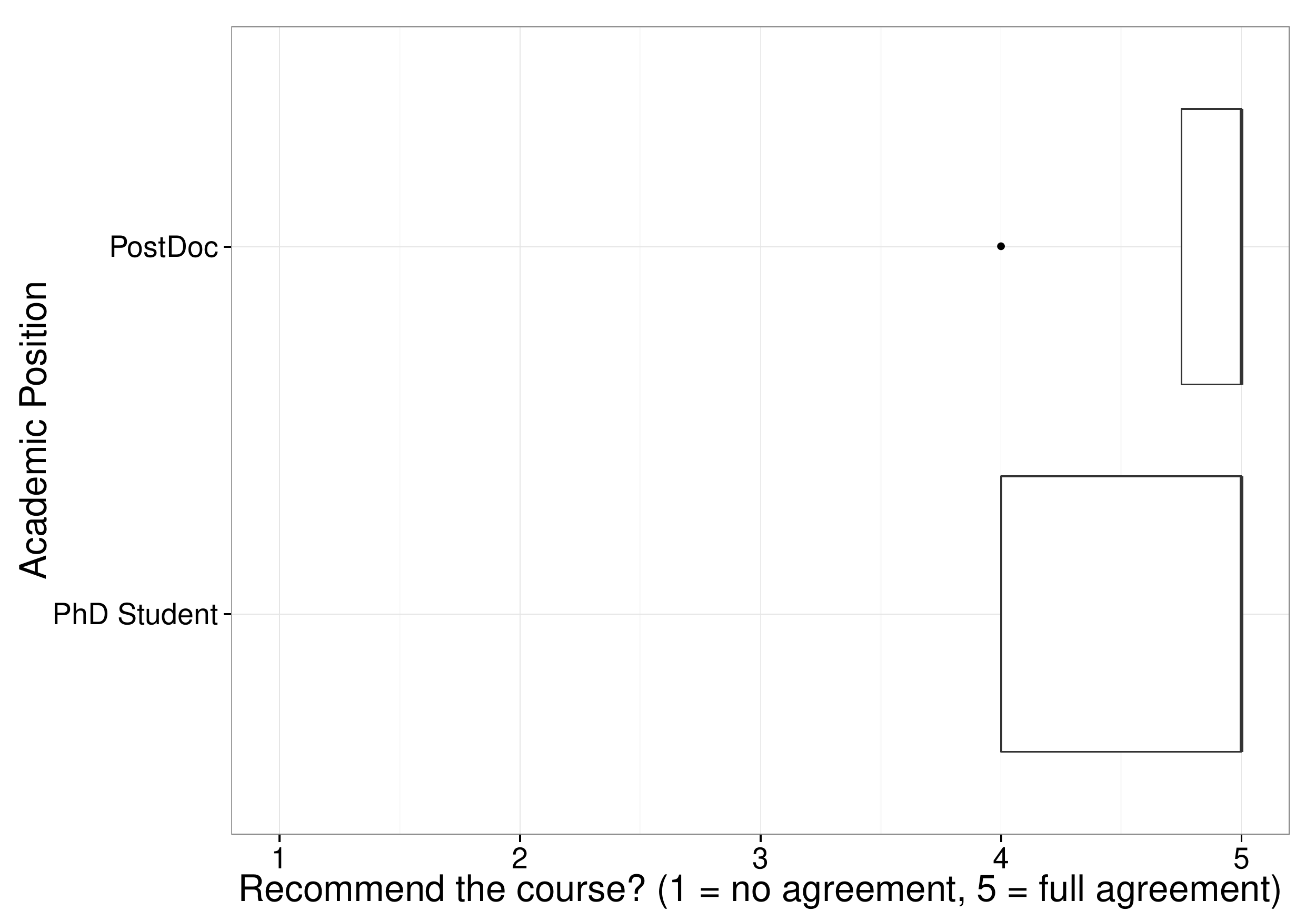}
    \caption{Candidates reply if they would recommend the workshop to others.}
    \label{sfig:ws_recommended}
  \end{subfigure}%

\caption{Box plots of the candidates estimates on whether they felt the APC contents to be relevant for their work (sub-figure \ref{sfig:ws_relevance} and whether participants would recommend the workshop to their peers (sub-figure \ref{sfig:ws_recommended}). The vertical box outer limits denoted the $25\%$ and $75\%$ quantile limits of the answer. The bold vertical line inside the box marks the arithmetic mean of the sample. }
\label{fig:ws_posteriori}
\end{figure}

Figure \ref{fig:ws_posteriori} illustrates the effect that the APC school had on its participants. 
Figure \ref{sfig:ws_relevance} indicates that the material conveyed is fully relevant 
for the work of the PhD students that came. 
The Post-Docs appear to consider it to be relevant, but not to $100\%$. 
This might be due to several factors: first, 
Post-Docs are not expected 
to spent all of their time dedicated to code. 
%Especially in the university environment, they have to contribute to teaching 
%and student supervision, budget acquisition etc. 
Second, PhD students are more likely to be in the situation of adding new feature sets 
to existing source code motivated by a physics question.
%as Post-Docs act more on the scientific stage rather than on a code base. \\

Figure \ref{sfig:ws_recommended} emphasizes that the course focus, 
teaching and overall layout of the APC was well received by all participants 
as both Post-Docs and PhD students would fully recommend the course to others. 

Although limited by the small data sample size at the time of submitting this paper, the 
statistical data analysis appears consistent with the immediate evaluation questionnaires and 
the informal feedback collected by 
some of the lecturers in their direct interactions with APC workshop participants.
It confirms that the need for training, mentoring and collaborative 
exercise of advanced software design concepts is very high and that 
the community would profit greatly from an extension or continuation of the efforts established in APC. \\